\documentclass[aps,prl,reprint,letterpaper,showpacs,twocolumn,nofootinbib]{revtex4-1}
\usepackage[UKenglish]{babel}
\usepackage{graphicx}
\usepackage[pdftex                     % using pdftex
           ,pagebackref=false          % put link back to original page ?
           ,colorlinks=false           % color link numbers
           ]{hyperref}
\hypersetup{linkcolor= blue,           % figures, equations, refs
            citecolor= red,            % citations
            urlcolor=black}            % URLs

\usepackage{amsmath}
\usepackage{mathtools}
\usepackage{eufrak}
\usepackage{times}

\newcommand*{\fleur}{\texttt{FLEUR}}
\newcommand*{\wprog}{\textsc{wannier}{\footnotesize{90}}}
\newcommand*{\DM}{Dzyaloshinskii-Moriya}
\newcommand*{\vn}{\boldsymbol}
\newcommand*{\I}{\mathrm{i}}

\setcitestyle{super,open={},close={}}
\renewcommand{\onlinecite}[1]{\nocite{#1}\citenum{#1}}
\makeatletter
\renewcommand\@biblabel[1]{#1.}
\makeatother

\begin{document}

%..............ABSTRACT.............
\begin{abstract}
{\bf Reliable and energy efficient magnetization switching by electrically-induced spin-orbit torques is of crucial technological relevance for spintronic devices implementing memory and logic functionality. Here we predict that the strength of spin-orbit torques and the related \DM{} interaction in topologically non-trivial magnetic insulators can exceed by far that of conventional metallic magnets. In analogy to the quantum anomalous Hall effect, we explain this extraordinary response in absence of longitudinal currents as a hallmark of magnetic monopoles in the electronic structure of systems that are interpreted most naturally within the framework of {\itshape mixed Weyl semimetals}. We thereby launch the effect of spin-orbit torque into
the field of topology and reveal its crucial role in mediating the topological phase transitions 
%Remarkably, the magnetization switching by anti-damping torques in mixed Weyl semimetals can induce topological phase transitions due to 
arising due to the complex interplay between magnetization direction and momentum-space topology. 
The concepts presented here may be exploited to understand and utilize magneto-electric coupling phenomena in insulating ferromagnets and antiferromagnets.}
\end{abstract}
%Universal concepts of topology and Berry phases revolutionized our understanding of various electronic properties in condensed-matter physics. are nowadays powerful tools in analyzing and predicting various electronic properties in condensed-matter physics.
%The universal concepts of topology and Berry phases became powerful tools in the analysis and prediction of various electronic properties of condensed-matter systems during the last decades. Generalizing these ideas, 

\setcounter{secnumdepth}{3}
%..............TITLE................
 \title{Mixed Weyl semimetals and dissipationless magnetization control in insulators by spin-orbit torques}
 \author{Jan-Philipp Hanke}
 \email{j.hanke@fz-juelich.de}
 \author{Frank Freimuth}
 \author{Chengwang Niu}
 \author{Stefan Bl\"ugel}
 \author{Yuriy Mokrousov}
 \date{April 24, 2017}
 %\pacs{XXX}
 \affiliation{Peter Gr\"unberg Institut and Institute for Advanced Simulation,\\Forschungszentrum J\"ulich and JARA, 52425 J\"ulich, Germany}
 \maketitle

Progress in control and manipulation of the magnetization in magnetic materials is pivotal for the innovative design of future non-volatile, high-speed, low power, and scalable spintronic devices. The effect of spin-orbit torque (SOT) provides an efficient means of magnetization control by electrical currents in systems that combine broken spatial inversion symmetry and spin-orbit interaction~\cite{Chernyshov2009,Miron2010,Miron2011,Garello2013,Freimuth2014a}. These current-induced torques are believed to play a key role in the practical implementation of various spintronics concepts, since they were demonstrated to mediate the switching of single ferromagnetic layers~\cite{Miron2011a,Liu2012} and antiferromagnets~\cite{Wadley2016} via the exchange of spin angular momentum between the crystal lattice and the (staggered) collinear magnetization. Among the two different contributions to SOTs, the so-called anti-damping torques are of utter importance owing to the robustness of their properties with respect to details of disorder~\cite{Freimuth2014a}.

Only recently, the research on electrically-controlled magnetization switching started to reach out to topological condensed matter $-$ for example, very efficient magnetization switching has been achieved lately in metallic systems incorporating topological insulators~\cite{Mellnik2014}. And although in latter cases a strong torque can be generated, the resulting electric-field response does not rely on the global topological properties of these trivial systems. The discovery of a quantized version of the anomalous Hall effect in magnetic insulators with non-trivial topology in momentum space~\cite{Haldane1988,Yu2010,Chang2013} led to a revolution in forging new spintronic device concepts that utilize topology. On the other hand, moving the field of magnetization control by SOTs into the realm of topological spintronics would open bright avenues in exploiting universal arguments of topology for designing magneto-electric coupling phenomena in magnetic insulators.
With this work, we firmly put the phenomenon of SOT on the topological ground. Employing theoretical techniques we investigate the origin and size of anti-damping SOTs and \DM{} interaction (DMI) in prototypes of topologically non-trivial magnetic insulators, demonstrate
that complex topological properties have a direct strong impact 
on the emergence and magnitude of SOT and DMI in various classes of
magnetic insulators, and formulate intriguing perspectives for the electric-field control of magnetization in absence of longitudinal charge currents. 
%: (i) magnetically doped graphene, and (ii) a functionalized bismuth film. 
%We demonstrate that these systems can be interpreted naturally as realizations of {\itshape mixed Weyl semimetals} in the composite phase space of crystal momentum and magnetization direction. Remarkably, we find that the magnitudes of SOTs and DMI in the considered family of insulators can reach gigantic values exceeding by far those of conventional metallic ferromagnets. We show that emergent magnetic monopoles are pivotal in giving rise to these  effects, and 
%We point out intriguing perspectives for the electric-field control of magnetization in absence of longitudinal charge currents in these systems, which may ultimately lead to an ultralow power consumption of future spintronic devices exploiting SOTs.

\begin{figure}%[b]
\centering
\includegraphics{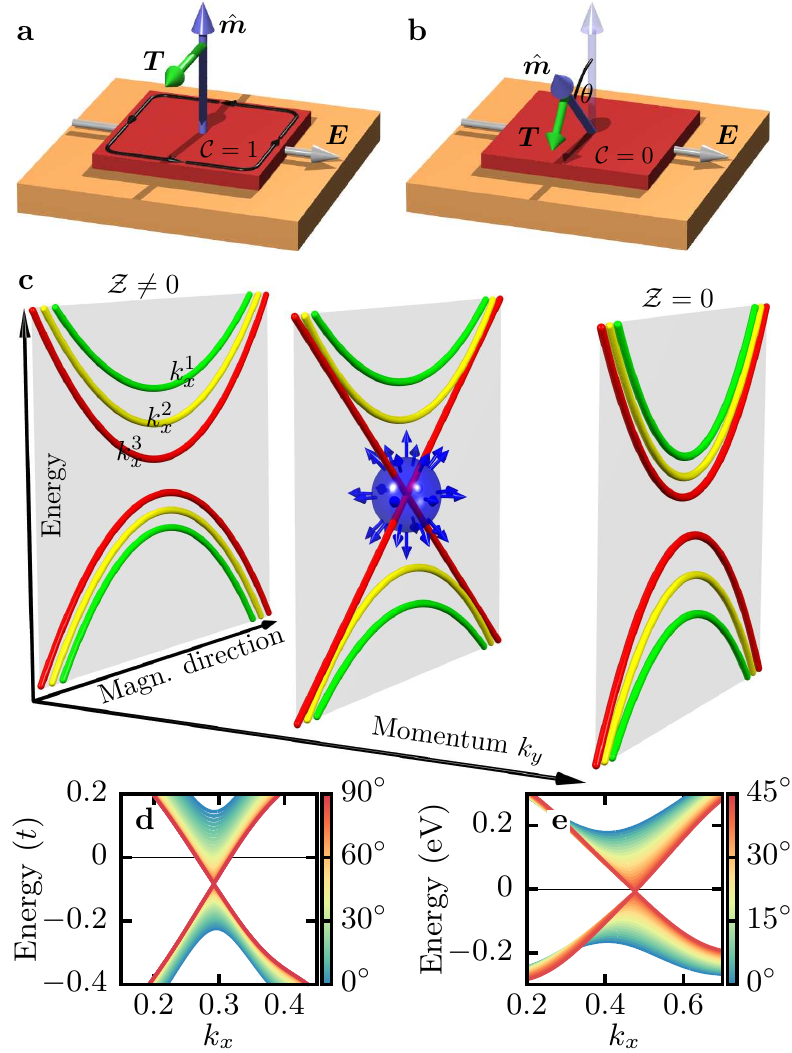}
\caption{{\bf Emergence of mixed Weyl points.} ({\bf a})~The magnetization $\hat{\vn m}$ of a topologically non-trivial insulator is subject to the anti-damping torque $\vn T$ if an electric field $\vn E$ is applied. ({\bf b})~The resulting reorientation of the magnetization by $\theta$ can trigger a topological phase transition to the trivial insulator. ({\bf c})~Schematic evolution of two energy bands in the complex phase space of crystal momentum and magnetization direction, where the colors of the bands indicate different $k_x$. If $k_y$ is tuned, the electronic structure displays a monopole, which is correlated with a change in the mixed Chern number $\mathcal Z$. Such crossing points are observed in ({\bf d})~the model of magnetically doped graphene with hopping $t$, and ({\bf e})~the functionalized bismuth film, where colors indicate the magnetization direction $\hat{\vn m}=(\sin\theta,0,\cos\theta)$. The shown monopoles arise at $\theta=90^\circ$ and $\vn k=(0.29\frac{2\pi}{a_x},0.41\frac{2\pi}{a_y})$ for ({\bf d}), and $\theta=43^\circ$ and $\vn k=(0.48,0.19)$ in internal units for ({\bf e}).}
\label{fig:monopole}
\end{figure}

\vspace{0.5cm}
\noindent{\bf Results}\\
{\bf Mixed Weyl semimetals and spin-orbit torque}. In a clean sample, the anti-damping SOT $\vn T$ acting on the magnetization in linear response to the electric field $\vn E$ is mediated by the so-called torkance tensor $\tau$, i.e., $\vn T=\tau \vn E$~\cite{Freimuth2014} (see Fig.~\ref{fig:monopole}a,b). The Berry phase nature of the anti-damping SOT manifests in the fact that the tensor elements $\tau_{ij}$ are proportional to the {\itshape mixed} Berry curvature $\Omega^{\hat{\vn m}\vn k}_{ij}= \hat{\vn e}_i \cdot 2\text{Im}\sum_n^{\text{occ}}\langle \partial_{\hat{\vn m}} u_{\vn k n}| \partial_{k_j} u_{\vn k n}\rangle$ of all occupied states~\cite{Freimuth2014,Kurebayashi2014}, which incorporates derivatives of lattice-periodic wave functions $u_{\vn k n}$ with respect to both crystal momentum $\vn k$ and magnetization direction $\hat{\vn m}$. Here, $\hat{\vn e}_i$ denotes the $i$th Cartesian unit vector. Intimately related to the anti-damping SOT is the DMI~\cite{Dzyaloshinsky1958,Moriya1960}, crucial for the emergence of chiral domain walls and chiral skyrmions~\cite{Neubauer2009,Kanazawa2011,Franz2014,Gayles2015}, which can be quantified by the so-called spiralization tensor $D$ reflecting the change of the free energy $F$ due to chiral perturbations $\partial_j\hat{\vn m}$ according to $F=\sum_{ij}D_{ij}\hat{\vn e}_i\cdot(\hat{\vn m}\times \partial_j \hat{\vn m})$~\cite{Freimuth2014}.

Optimizing the efficiency of magnetization switching in spintronic devices by current-induced SOTs relies crucially on the knowledge of the microscopic origin of most prominent contributions to the electric-field response. To promote the understanding, it is rewarding to draw an analogy between the anti-damping SOT as given by $\Omega^{\hat{\vn m}\vn k}_{ij}$ and the intrinsic anomalous Hall effect as determined by the Berry curvature $\Omega^{\vn k \vn k}_{ij}=2\text{Im}\sum_n^{\text{occ}}\langle\partial_{k_i} u_{\vn k n}|\partial_{k_j} u_{\vn k n}\rangle$~\cite{Nagaosa2010}. Both $\Omega^{\vn k \vn k}$ and $\Omega^{\hat{\vn m}\vn k}$ are components of a general curvature tensor $\vn\Omega$ in the composite $(\vn k,\hat{\vn m})$ phase space combining crystal momentum and magnetization direction~\cite{Xiao2005,Freimuth2013}. Band crossings, also referred to as magnetic monopoles in $\vn k$-space, are known~\cite{Fang2003} to act as important sources or sinks of $\Omega^{\vn k \vn k}$. When transferring this concept to current-induced torques, crossing points in the composite phase space can be anticipated to give rise to a large mixed Berry curvature $\Omega^{\hat{\vn m}\vn k}$, which in turn yields the dominant microscopic contribution to torkance and spiralization. Thus, materials providing such monopoles close to the Fermi energy can be expected to exhibit notably strong SOTs and DMI.

\begin{figure}
\centering
\includegraphics{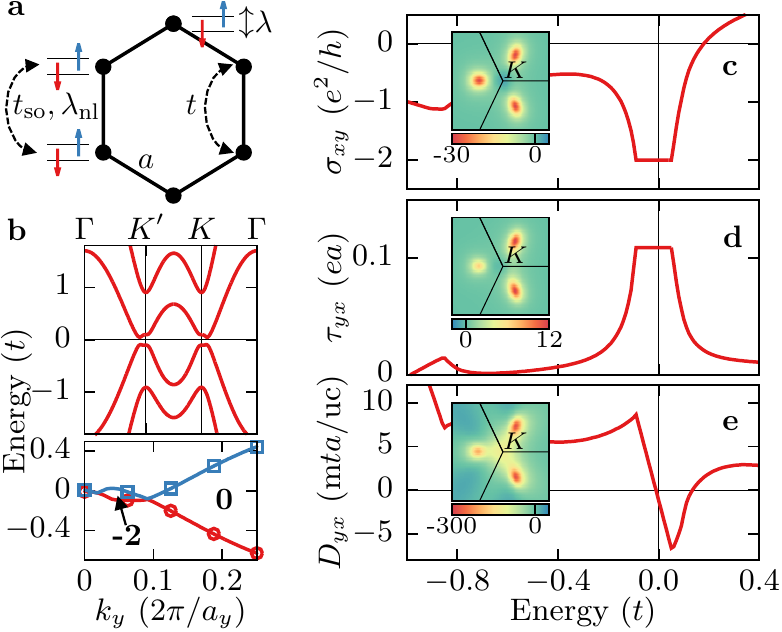}
\caption{{\bf Model of magnetically doped graphene.} ({\bf a})~Sketch of the tight-binding model. ({\bf b})~Top: Band structure with out-of-plane magnetization and $t_\text{so}=0.3t$, $\lambda=0.1t$, $\lambda_\text{nl}=0.4t$. Bottom: Valence band maximum (red circles) and conduction band minimum (blue squares) in the $(k_x,\theta)$-space. The bold integers denote the mixed Chern number $\mathcal Z$ in the insulating regions, and $a_y=3a/2$. ({\bf c})--({\bf e}) Energy dependence of the anomalous Hall conductivity $\sigma_{xy}=\mathcal C e^2/h = e^2/(2\pi h)\int\Omega^{\vn k\vn k}_{xy}dk_x dk_y$, the torkance $\tau_{yx}$, and the spiralization $D_{yx}$, respectively, for an out-of-plane magnetization. Insets show the corresponding momentum-space distributions summed over all occupied states in the vicinity of the $K$-point.}
\label{fig:model}
\end{figure}

In the field of topological condensed matter~\cite{Hasan2010,Qi2011}, the recent advances in the realization of quantum anomalous Hall, or, Chern insulators have been striking~\cite{Yu2010,Chang2013}. These magnetic materials are characterized by a quantized value of the anomalous Hall conductivity and an integer non-zero value of the Chern number in $\vn k$-space, $\mathcal C = 1/(2\pi)\int\Omega^{\vn k \vn k}_{xy}dk_x dk_y$. On the other hand, topological semimetals have recently attracted great attention due to their exceptional properties stemming from monopoles in momentum space. Among these latter systems, magnetic Weyl semimetals  host gapless low-energy excitations with linear dispersion in the vicinity of non-degenerate band crossings at generic $\vn k$-points~\cite{Xu2011,Xu2015,Lv2015,Gosalbez2015}, which are sources of $\Omega^{\vn k \vn k}$. Their conventional description in terms of the Weyl Hamiltonian can be formally extended to the case of what we call the {\itshape mixed Weyl semimetal} as described by $H_W=v_x k_x \sigma_x + v_y k_y \sigma_y + v_{\theta} \theta \sigma_z$, where $\vn \sigma=(\sigma_x,\sigma_y,\sigma_z)$ is the vector of Pauli matrices, and $\theta$ is the angle that the magnetization $\hat{\vn m}=(\sin\theta,0,\cos\theta)$ makes with the $z$-axis. As illustrated in Fig.~\ref{fig:monopole}c, mixed Weyl semimetals feature monopoles in the composite phase space of $\vn k$ and $\theta$, which are sources of the general curvature $\vn\Omega$. In analogy to conventional Weyl semimetals~\cite{Xu2011}, we can characterize the topology and detect magnetic monopoles by monitoring the flux of the mixed Berry curvature through planes of constant $k_y$ as given by the integer {\it mixed} Chern number $\mathcal Z=1/(2\pi)\int \Omega^{\hat{\vn m}\vn k}_{yx}d\theta dk_x$,~Fig.~\ref{fig:monopole}c. In the following, we show that a significant electric-field response near monopoles in mixed Weyl semimetals is invaluable in paving the road towards dissipationless magnetization control by SOTs~\cite{Avci2016}.  
\\
{\bf Magnetically doped graphene.} We begin with a tight-binding model of magnetically doped graphene~\cite{Qiao2010}:
\begin{equation}
\begin{split}
H &= -t \sum\limits_{\langle ij \rangle\alpha}  c_{i\alpha}^\dagger c_{j\alpha}^{\phantom{\dagger}} + \I t_\text{so}\sum\limits_{\langle ij \rangle\alpha \beta}  \hat{\vn e}_z \cdot (\vn\sigma \times {\vn d}_{ij}) c_{i\alpha}^\dagger c_{j\beta}^{\phantom{\dagger}}\\
&+ \lambda \sum_{i\alpha \beta} (\hat{\vn m}\cdot \vn\sigma) c_{i\alpha}^\dagger c_{i\beta}^{\phantom{\dagger}} - \lambda_\text{nl} \sum\limits_{\langle ij \rangle\alpha \beta}  (\hat{\vn m}\cdot \vn \sigma) c_{i\alpha}^\dagger c_{j\beta}^{\phantom{\dagger}}\, ,
\end{split}
\label{eq:model}
\end{equation}
which is sketched in Fig.~\ref{fig:model}a. Here, $c_{i\alpha}^\dagger$ ($c_{i\alpha}^{\vphantom{\dagger}}$) denotes the creation (annihilation) of an electron with spin $\alpha$ at site $i$, $\langle ...\rangle$ restricts the sums to nearest neighbors, and the unit vector $\vn d_{ij}$ points from $j$ to $i$. Besides the usual %nearest-neighbor
hopping with amplitude $t$, the first line in equation~\eqref{eq:model} contains the Rashba spin-orbit coupling of strength $t_\text{so}$ originating in the surface potential gradient of the substrate. The remaining terms in equation~\eqref{eq:model} are the exchange energy due to the local ($\lambda$) and non-local ($\lambda_\text{nl}$) exchange interaction between spin and magnetization. Depending on $\hat{\vn m}$, the non-local exchange describes a hopping process during which the spin can flip. Supplementary Note 1 provides further details on the tight-binding model and its numerical solution.

First, by monitoring the evolution of the mixed Chern number $\mathcal Z$ we demonstrate that the above model hosts a mixed Weyl semimetal state. Indeed, as shown in Fig.~\ref{fig:model}b, the topological index $\mathcal Z$ changes from $-2$ to $0$ at a certain value of $k_y$, indicating thus the presence of a band crossing in composite phase space that carries a topological charge of $+2$. One of these monopoles appears near the $K$-point off any high-symmetry line if the magnetization is oriented in-plane along the $x$-direction (see Fig.~\ref{fig:monopole}d). The emergence of the quantum anomalous Hall effect~\cite{Qiao2010}, Fig.~\ref{fig:model}c, over a wide range of magnetization directions can be understood as a direct consequence of the magnetic monopoles acting as sources of the curvature $\Omega^{\vn k \vn k}$. Correspondingly, for $\hat{\vn m}$ out of the plane, the system is a quantum anomalous Hall insulator. Moreover, large values of the mixed curvature $\Omega^{\hat{\vn m}\vn k}$ in the vicinity of the monopole are visible in the momentum-space distributions of torkance and spiralization in the insets of Figs.~\ref{fig:model}d and~\ref{fig:model}e, respectively. For an out-of-plane magnetization, the primary microscopic contribution to the effects arises from an avoided crossing along $\Gamma K$ -- a residue of the Weyl point in $(\vn k,\theta)$-space. Since the expression
for the mixed Berry curvature relies only on the derivative of the wavefunction with respect to one of the components of the Bloch vector, 
%the expressions~\eqref{eq:torkance} and~\eqref{eq:spiralization} differentiate the wave function only with respect to one of the momentum coordinates, 
the symmetry between $k_x$ and $k_y$ in the distributions of torkance and spiralization is broken naturally (see Methods).

As a consequence of the monopole-driven momentum-space distribution, the energy dependence of the torkance $\tau_{yx}$, Fig.~\ref{fig:model}d, displays a decent magnitude of $0.1\,ea$ in the insulating region (with $a$ being the interatomic distance), and stays constant throughout the band gap. In contrast to the Chern numbers $\mathcal C$ and $\mathcal Z$, the torkance $\tau_{yx}$ is, however, not guaranteed to be quantized to a robust value,~i.e., the height of the torkance plateau in Fig.~\ref{fig:model}d is sensitive to fine details of the electronic structure such as magnetization direction and model parameters. Because of their intimate relation in the Berry phase theory~\cite{Freimuth2014,Thonhauser2011,Hanke2016}, the plateau in torkance implies a linear behavior of the spiralization $D_{yx}$ within the gap, changing from $8\,$m$ta$/uc to $-6\,$m$ta$/uc as shown in Fig.~\ref{fig:model}e, where ``uc" refers to the in-plane unit cell containing two atoms.

\begin{figure*}
\centering
%\includegraphics{fig3a_draft}
%\hspace{0.9cm}
%\includegraphics{fig3b_draft}
\includegraphics{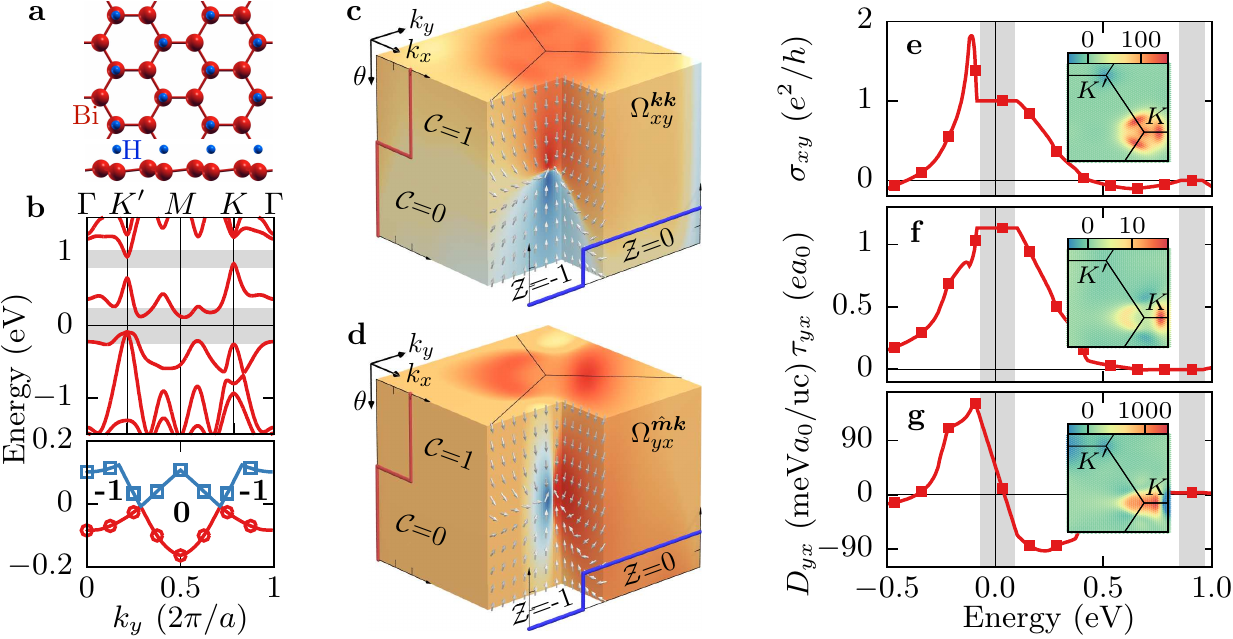}
\caption{{\bf Electronic structure and magneto-electric properties of a mixed Weyl semimetal.} ({\bf a})~Crystal structure of the semi-hydrogenated Bi(111) bilayer. ({\bf b})~Top: First-principles band structure for an out-of-plane magnetization, where the region of the topologically complex band gap and the trivial one above are highlighted. Bottom: Evolution of valence band maximum (red circles) and conduction band minimum (blue squares) in the $(k_x,\theta)$-space. Bold integers denote the mixed Chern number $\mathcal Z$, and $a$ is the in-plane lattice constant. ({\bf c})--({\bf d})~Monopole-like field of momentum space and mixed Berry curvatures near one of the mixed Weyl points. Arrows indicate the direction of the curvature field $(-\Omega^{\hat{\vn m}\vn k}_{yy}, \Omega^{\hat{\vn m}\vn k}_{yx}, \Omega^{\vn k \vn k}_{xy})$, and a logarithmic color scale is used to display two of its components, where dark red (dark blue) denotes large positive (negative) values. ({\bf e})--({\bf g})~Energy dependence of 
$\sigma_{xy}$, $\tau_{yx}$, and $D_{yx}$ for magnetization perpendicular to the film plane. Insets show the microscopic distributions in momentum space near $K$ and $K^\prime$.}
\label{fig:bih}
\end{figure*}

\begin{figure*} %[b]
\centering
\includegraphics{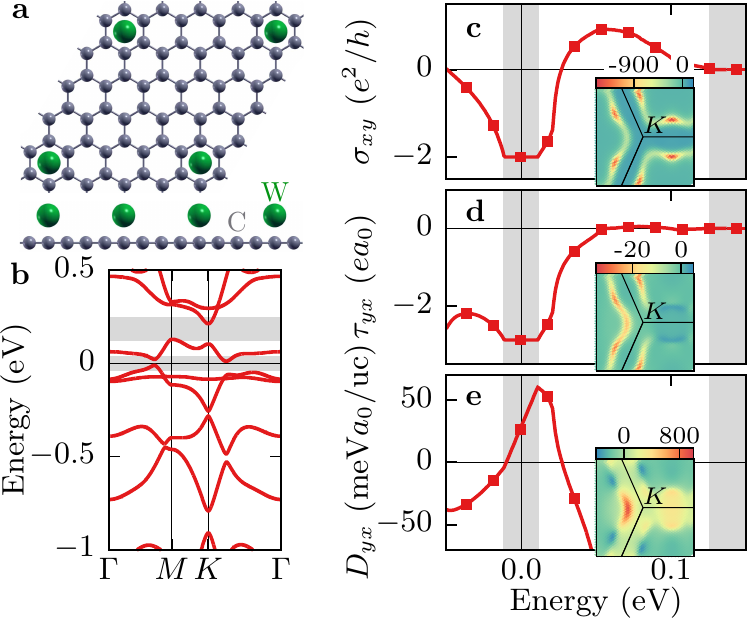}
\hspace{0.9cm}
\includegraphics{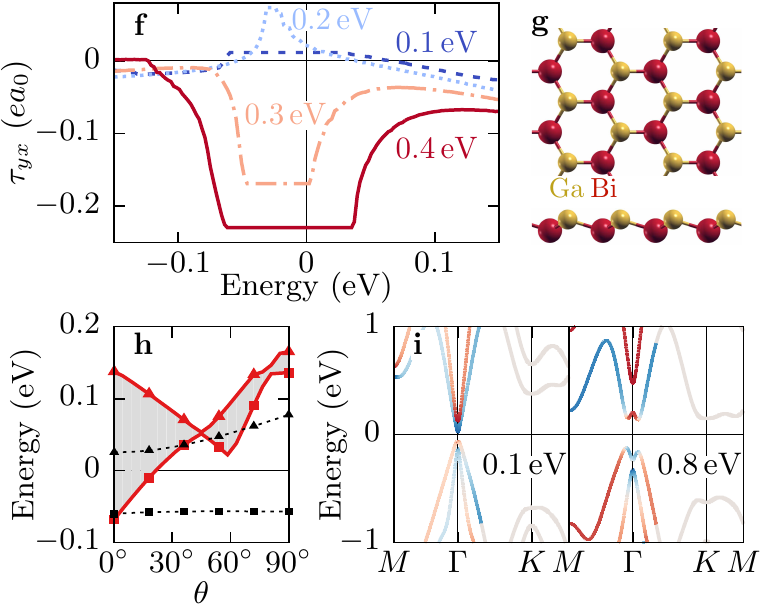}
\caption{{\bf Monopole-driven spin-orbit torques in mixed Weyl semimetals.} ({\bf a})~Crystal structure of graphene decorated by W adatoms in $4$$\times$$4$ geometry. ({\bf b})~First-principles band structure for an out-of-plane magnetization. The topologically non-trivial gap around the Fermi level and the trivial gap above are highlighted. ({\bf c})--({\bf e})~Energy dependence of anomalous Hall conductivity $\sigma_{xy}$, torkance $\tau_{yx}$, and spiralization $D_{yx}$, respectively. Insets show the microscopic distributions in momentum space near the $K$-point. ({\bf f})~Energy dependence of the torkance $\tau_{yx}$ in a GaBi film upon applying an exchange field $\vn B=B_0(\sin \theta,0,\cos\theta)$ perpendicular to the film plane, i.e., $\theta=0^\circ$. Numbers denote the value of $B_0$. ({\bf g})~Crystal structure of the system. ({\bf h})~Evolution of valence band maximum (squares) and conduction band minimum (triangles) with $\theta$ for $B_0=0.1\,$eV (dashed black) and $B_0=0.8\,$eV (solid red). ({\bf i})~Band structures for $\theta=0^\circ$ and two different values of $B_0$, where colors encode the spin polarization perpendicular to the film.}
\label{fig:wgr_gabi}
\end{figure*}

To provide a realistic manifestation of the model considerations above, we study from {\it ab initio} systems of graphene decorated by transition-metal adatoms, Fig.~\ref{fig:wgr_gabi}a. These systems, which exhibit complex spin-orbit mediated hybridization of graphene $p$ states with $d$ states of the transition metal, have by now become one of the prototypical material classes for realization of the quantum anomalous Hall effect~\cite{Ding2011,Zhang2012,Acosta2014,Hu2015,Ren2016}. Details on the first-principles calculations are provided in Supplementary Note 2. In the Chern insulator phase of these materials with magnetization perpendicular to the graphene plane, depending on the transition-metal adatom, both torkance and spiralization can reach colossal magnitudes that originate from mixed Weyl points. In the case of W in $4$$\times$$4$-geometry on graphene, for example, the torkance amounts to a huge value of $\tau_{yx}=-2.9\, ea_0$ (with $a_0$ being Bohr's radius), and the spiralization $D_{yx}$ ranges from $-5\,$meV$a_0$/uc to $60\,$meV$a_0$/uc, Fig.~\ref{fig:wgr_gabi}b-e, surpassing thoroughly the values obtained in metallic magnetic heterostructures~\cite{Freimuth2014a,Freimuth2014} and non-centrosymmetric bulk magnets~\cite{Gayles2015}. Since the details of the electronic structure can influence the value of the torkance in the gap, upon replacing W with other  transition metals, the magnitude of SOT and DMI can be tailored in the gapped regions of corresponding materials according to our calculations.
\\
{\bf Functionalized bismuth film.} Aiming at revealing pronounced magneto-electric coupling effects in magnetic insulators with larger band gaps as compared to the above examples, we turn to a semi-hydrogenated Bi(111) bilayer, Fig.~\ref{fig:bih}a, which is a prominent example of functionalized
insulators realizing non-trivial topological phases~\cite{Ren2016}.
As we show, semi-hydrogenated Bi(111) bilayer is a mixed Weyl semimetal.  For an out-of-plane magnetization, the system is a valley-polarized quantum anomalous Hall insulator~\cite{Niu2015} with a magnetic moment of $1.0\,\mu_B$ per unit cell, and it exhibits a large band gap of $0.18\,$eV at the Fermi energy as well as a distinct asymmetry between the valleys $K$ and $K^\prime$, Fig.~\ref{fig:bih}b.

Analyzing the evolution of the mixed Chern number $\mathcal Z$ as a function of $k_y$ in Fig.~\ref{fig:bih}b, we detect two magnetic monopoles of opposite charge that emerge at the transition points between the topologically distinct phases with $\mathcal Z=-1$ and $\mathcal Z=0$. Alternatively, these crossing points and the monopole charges in the composite phase space could be identified by monitoring the variation of the momentum-space Chern number $\mathcal C$ with magnetization direction. These monopoles occur at generic points near the valley $K$ for $\theta=43^\circ$ (see Fig.~\ref{fig:monopole}e) and in the vicinity of the $K^\prime$-point for $\theta= 137^\circ$, respectively. The presence of such mixed Weyl points in the electronic structure drastically modifies the behavior of the general curvature $\vn \Omega$ in their vicinity, as visible from the  three-dimensional representation of $\vn \Omega$ displayed in Fig.~\ref{fig:bih}c,d. Revealing characteristic sign changes when passing through monopoles in composite phase space, the singular behavior of the Berry curvature underlines the role of the mixed Weyl points as sources or sinks of  $\vn \Omega$. For an out-of-plane magnetization, the complex nature of the electronic structure in momentum space manifests in the quantization of $\mathcal{C}$ to $+1$, Fig.~\ref{fig:bih}e, which is primarily due to the pronounced positive contributions near $K$. Calculations of the energy dependence of the 
torkance and spiralization in the system, shown in Figs.~\ref{fig:bih}f and~\ref{fig:bih}g, reveal the extraordinary magnitudes of these phenomena of the order of $1.1\,ea_0$ for $\tau_{yx}$ and $50\,$meV$a_0$/uc for $D_{yx}$, exceeding by far the typical magnitudes of these effects in magnetic metallic materials~\cite{Freimuth2014a,Freimuth2014,Gayles2015}.
\\
{\bf Proof of monopole-driven SOT enhancement.}
An important question to ask at this point is whether the colossal magnitude of the SOT in the insulators considered above can be unambiguously identified with the mixed Weyl semimetallic state. In the following, we answer this question by explicitly demonstrating the utter importance of the emergent mixed monopoles for driving pronounced magneto-electric response. First, by removing the mixed Weyl points from the electronic structure of the model~\eqref{eq:model} via, e.g., including an intrinsic spin-orbit coupling term, we confirm that the electric-field response is strongly suppressed, which promotes the monopoles as unique origin of large SOT and DMI. Secondly, to verify this statement from the first-principles calculations, we analyze the electric-field response throughout the topologically trivial gaps above the Fermi level that are highlighted in Figs.~\ref{fig:bih}b and~\ref{fig:wgr_gabi}b. Since these gaps do not exhibit the mixed Weyl points, we obtain a greatly diminished magnitude of the torkance $\tau_{yx}$ within these energy regions as apparent from Figs.~\ref{fig:bih}f and~\ref{fig:wgr_gabi}d.

Finally, we clearly demonstrate the key role of these special points by studying an illustrative example: a thin film of GaBi with triangular lattice structure, Fig.~\ref{fig:wgr_gabi}g. The initial system is a non-magnetic trivial insulator, on top of which we artificially apply an exchange field $\vn B=B_0(\sin\theta,0,\cos\theta)$, with the purpose of triggering a topological
phase transition as a function of the exchange field strength, see Supplementary Note 4. When tuning the exchange field strength $B_0$ we carefully monitor the evolution of the system from a trivial magnetic insulator for $|B_0|\leq 0.2\,$eV to a mixed Weyl semimetal as indicated by the emergence of magnetic monopoles in the electronic structure. The latter phase is accompanied by the quantum anomalous Hall effect prominent for a finite range of directions $\theta$, for instance, if $\vn B$ is perpendicular to the film plane, Fig.~\ref{fig:wgr_gabi}h,i. Comparing in Fig.~\ref{fig:wgr_gabi}f the electric-field response for these two distinct phases, we uniquely identify drastic changes in sign and magnitude of the torkance $\tau_{yx}$ with the transition from the trivial insulator to the mixed Weyl semimetal hosting monopoles near the $\Gamma$-point. This proves the crucial relevance of emergent monopoles in driving magneto-electric coupling effects in topologically non-trivial magnetic insulators. %we note that the appearance of magnetic monopoles near the $\Gamma$-point drives a sign change of the torkance $\tau_{yx}$ upon entering the mixed Weyl semimetallic regime.

\vspace{0.5cm}
\noindent{\bf Discussion}
\\
Remarkably, the magnetization switching via anti-damping torques in mixed Weyl semimetals can be utilized to induce topological phase transitions from a Chern insulator to a trivial magnetic insulator mediated by the complex interplay between magnetization direction and momentum-space topology in these systems as illustrated in Fig.~\ref{fig:monopole}a,b. In the case of the functionalized bismuth film, for instance, the material is a trivial magnetic insulator with a band gap of $0.25\,$eV if the magnetization is oriented parallel to the film plane. Nevertheless, the resulting anti-damping torkance in this trivial state
%$\tau_{yx}=0.5\,ea_0$ 
is still very large, and the DMI exhibits a strong variation 
%varies from $D_{yx}=105\,$meV$a_0$/uc to $-60\,$meV$a_0$/uc 
within the gap, see Supplementary Note 3. 
We therefore motivate experimental search and realization of large magneto-electric response and topological phase transitions in quantum anomalous Hall systems fabricated to date~\cite{Chang2013,Checkelsky2014,Kou2014,Chang2015}. 
Overall, mixed Weyl semimetals that combine  exceptional electric-field response with a large band gap (such as, e.g., functionalized bismuth films) lay out extremely promising vistas in room-temperature applications of magneto-electric coupling phenomena for dissipationless magnetization control -- a subject which is currently under extensive scrutiny (see, e.g., refs.~\onlinecite{Avci2016,Chu2008,Chiba2008}).
In contrast to the anti-damping SOT in magnetic metallic bilayers (such as Co/Pt) for which large spin-orbit interaction
in the non-magnetic substrate is necessary for generating large spin Hall effect and large values of SOT~\cite{Garello2013}, the 
magnitude of the SOT in insulating phases of a mixed Weyl semimetal is driven by the presence of the mixed monopole rather than the spin-orbit strength itself. This opens perspectives in exploiting a strong magneto-electric response of weak-spin-orbit materials.  

%Toward this goal, also elements with moderate spin-orbit interactions provide fruitful prospects since the predicted enhancement of magneto-electric phenomena roots not in the spin-orbit interaction but is driven entirely by the emergence of mixed Weyl points in the spectrum.

In the examples that we considered here, the non-trivial topology of mixed Weyl semimetals leads to DMI changes over a wide range of values throughout the bulk band gap, implying that proper electronic-structure engineering enables us to tailor both strength and sign of the DMI in a given system, for instance, by doping or applying strain. Such versatility could be particularly valuable for the stabilization of chiral magnetic structures such as skyrmions in insulating ferromagnets. In the latter case, very large values of the anti-damping SOT arising in these systems would open exciting perspectives in manipulation and dynamical properties of chiral objects associated with minimal energy consumption by magneto-electric coupling effects. Generally, we would like to remark that magnetic monopoles in the composite phase space, which we discuss here, do not only govern the electric-field response in insulating magnets but are also relevant in metals, where they appear on the background of metallic bands.
Ultimately, in analogy to the (non-quantized) anomalous Hall effect in metals, this makes the analysis of SOT and DMI in metallic systems very complex owing to competing contributions to these effects from various bands present at the Fermi energy. In addition, the electric-field strength in metals is typically much smaller, limiting thus the reachable magnitude of response phenomena as compared to insulators.

At the end, we reveal the relevance of the physics discussed here for antiferromagnets (AFMs) that satisfy the combined symmetry of time reversal and spatial inversion. SOTs in such antiferromagnets are intimately linked with the physics of Dirac fermions, which are doubly-degenerate elementary excitations with linear dispersion~\cite{Tang2016,Smejkal2017}. In these systems, the reliable switching of the staggered magnetization by means of current-induced torques has been demonstrated very recently~\cite{Wadley2016}. In analogy to the concept of mixed Weyl semimetals presented here, we expect that the notion of {\itshape mixed Dirac semimetals} in a combined phase space of crystal momentum and direction of the staggered magnetization vector will prove fruitful in understanding the microscopic origin of SOTs in insulating antiferromagnets. Following the very same interpretation that we formulated here for ferromagnets, monopoles in the electronic structure of AFMs can be anticipated to constitute prominent sources or sinks of the corresponding general non-Abelian Berry curvature, whose mixed band-diagonal components correspond to the sublattice-dependent anti-damping SOT, in analogy to the spin Berry curvature for quantum spin Hall insulators and Dirac semimetals~\cite{Murakami2007,Murakami2008,Yang2014}. Correspondingly, exploiting the principles of electronic-structure engineering for topological properties depending on the staggered magnetization could result in an advanced understanding and utilization of pronounced magneto-electric response in insulating AFMs.

\vspace{0.5cm}
\noindent{\bf Methods}
\\
{\bf Tight-binding calculations.} The Hamiltonian~\eqref{eq:model} is a generalization of the model in ref.~\onlinecite{Qiao2010}, taking additionally into consideration arbitrary magnetization directions $\hat{\vn m}$ as well as the non-local exchange interaction. A brief description of its numerical solution is given in Supplementary Note 1.
\\
{\bf First-principles electronic structure calculations.} Using the full-potential linearized augmented plane-wave code \texttt{FLEUR}~\cite{fleur}, we performed self-consistent density functional theory calculations of the electronic structure of the considered materials using the structural parameters of refs.~\onlinecite{Zhang2012} and~\onlinecite{Niu2015}. The effect of spin-orbit coupling was treated within the perturbative second-variation scheme. Starting from the converged charge density, we constructed higher-dimensional Wannier functions~\cite{Hanke2015} by employing our extension of the \wprog{} code~\cite{Mostofi2014}. We used these functions to generalize the Wannier interpolation~\cite{Wang2006,Yates2007,Hanke2015} allowing us to evaluate efficiently anomalous Hall conductivity, torkance, and spiralization. Further details on the electronic structure calculations are given in Supplementary Note 2.
\\
{\bf Berry phase expressions for torkance and spiralization.}
In order to characterize the anti-damping SOTs, we evaluate within linear response the torkance~\cite{Freimuth2014}
\begin{equation}
\tau_{ij} = \frac{2e}{N_{\vn k}} \hat{\vn e}_i \cdot \sum\limits_{\vn k n}^{\text{occ}} \left[\hat{\vn m}\times \text{Im}\langle \partial_{\hat{\vn m}} u_{\vn k n} | \partial_{k_j} u_{\vn k n}\rangle\right] \, ,
\label{eq:torkance}
\end{equation}
where $N_{\vn k}$ is the number of $\vn k$-points, and $e>0$ denotes the elementary positive charge. Similarly, the spiralization~\cite{Freimuth2014} is obtained as
\begin{equation}
D_{ij} = \frac{\hat{\vn e}_i}{N_{\vn k}V}\cdot \sum\limits_{\vn kn}^{\text{occ}} \left[ \hat{\vn m} \times\text{Im} \langle \partial_{\hat{\vn m}} u_{\vn k n} | h_{\vn k n} | \partial_{k_j} u_{\vn k n}\rangle\right] \, ,
 \label{eq:spiralization}
\end{equation}
where $h_{\vn kn}=H_{\vn k}+\mathcal E_{\vn k n}-2\mathcal E_F$, $H_{\vn k}$ is the lattice-periodic Hamiltonian with eigenenergies $\mathcal E_{\vn k n}$, $\mathcal E_F$ is the Fermi level, and $V$ is the unit cell volume.
\\
{\bf Code availability.} The tight-binding code that supports the findings of this study is available from the corresponding authors on request.
\\
{\bf Data availability.} The data that support the findings of this study are available from the corresponding authors on request.

\vspace{0.5cm}
\def\bibsection{\noindent{\bf References}}
%\bibliographystyle{naturemag}
%\bibliography{my_bibliography}

\vspace{0.5cm}
\noindent{\bf Acknowledgements}
\\
We gratefully acknowledge computing time on the supercomputers JUQUEEN and JURECA at
J\"ulich Supercomputing Center as well as at the JARA-HPC cluster of RWTH Aachen, and funding under the HGF-YIG programme VH-NG-513 and SPP 1538 of DFG.

\vspace{0.5cm}
\noindent{\bf Author contributions}
\\
J.-P.H. uncovered the mixed Weyl points as origin of large magneto-electric coupling effects through model considerations and first-principles calculations. J.-P.H. and Y.M. wrote the manuscript. All authors discussed the results and reviewed the manuscript.

\vspace{0.5cm}
\noindent{\bf Additional information}
\\
{\bf Competing financial interests.} The authors declare no competing financial interests.

\newpage
\onecolumngrid
 \begin{center}
 \textbf{Supplementary Material: Mixed Weyl semimetals and dissipationless magnetization control in insulators by spin-orbit torques}\\ \vspace{0.3cm}
 Jan-Philipp Hanke, Frank Freimuth, Chengwang Niu, Stefan Bl\"ugel, and Yuriy Mokrousov\\
 \it{Peter Gr\"unberg Institut and Institute for Advanced Simulation,\\
 Forschungszentrum J\"ulich and JARA, 52425 J\"ulich, Germany}
 \end{center}

\makeatletter
\renewcommand{\fnum@figure}{Supplementary Figure \thefigure}
\makeatother
\setcounter{figure}{0}

\vspace{0.3cm}
\noindent{\bf Supplementary Note 1 | Tight-binding model}
\\
To arrive at the model Hamiltonian of the main text, the model in ref.~\onlinecite{Qiao2010} has been generalized to account for arbitrary magnetization directions $\hat{\vn m}$ and the non-local exchange interaction. We obtained a $4$$\times$$4$-matrix representation of the resulting Hamiltonian on the bipartite lattice of graphene by introducing four orthonormal basis states ${|N \alpha\rangle}$ that describe electrons with spin $\alpha=\{\uparrow,\downarrow\}$ on the sublattice $N=\{A,B\}$. Using Fourier transformations, we transformed this matrix to a representation $H(\vn k)$ in momentum space, which was subsequently diagonalized at every $\vn k$-point to access the electronic and topological properties of the system. The model parameters $t_\text{so}=0.3t$, $\lambda=0.1t$, and $\lambda_\text{nl}=0.4t$ were employed in this work. We chose the magnetization direction as $\hat{\vn m}=(\sin \theta,0,\cos\theta)$ for a direct comparison between the model and the first-principles calculations.
%Using the velocity operator $\vn v(\vn k)=\partial_{\vn k} H(\vn k)$ and the torque operator $\vn{\mathcal T} =  \hat{\vn m} \times \partial_{\hat{\vn m}} H(\vn k)$, we rewrote the wave function derivatives in Eqs.~\eqref{eq:ahc}--\eqref{eq:dmi} to compute anomalous Hall conductivity, torkance, and spiralization.
%The momentum-space integrals were converged using a uniform mesh of $512\times 512$ $\vn k$-points. Integrating the mixed Berry curvature over $XX$ $k_x$-points and $XX$ angles $\theta$, the mixed Chern number $\mathcal Z(k_y)$ was obtained.

\vspace{0.7cm}
\noindent{\bf Supplementary Note 2 | First-principles electronic structure calculations}
\\
Using the full-potential linearized augmented plane-wave code \texttt{FLEUR}~\cite{fleur}, we performed self-consistent density functional theory calculations of the electronic structure of (i) graphene decorated with W adatoms in $4$$\times$$4$-geometry, and (ii) a semi-hydrogenated Bi(111) bilayer. The structural and computational parameters of refs.~\onlinecite{Zhang2012} and~\onlinecite{Niu2015} were assumed in the respective cases. Starting from the converged charge density, the Kohn-Sham equations were solved on an equidistant mesh of $8$$\times$$8$ $\vn k$-points [$6$$\times$$6$ in case (i)] for $8$ different magnetization directions $\hat{\vn m}=(\sin\theta,0,\cos\theta)$, where the angle $\theta$ covers the unit circle once. Based on the resulting wave-function information in the composite phase space, we constructed a single set of higher-dimensional Wannier functions~\cite{Hanke2015} (HDWFs) for each of the systems by employing our extension of the \wprog{} code~\cite{Mostofi2014}. In case (i), we generated $274$ HDWFs out of $360$ energy bands with the frozen window up to $4\,$eV above the Fermi level, and in the case (ii), we extracted from $28$ bands $14$ HDWFs for a frozen window that extends to $2\,$eV above the Fermi energy.

We used the Wannier interpolation~\cite{Wang2006,Yates2007} that we generalized to treat crystal momentum and magnetization direction on an equal footing~\cite{Hanke2015} in order to evaluate the Berry curvatures $\Omega^{\vn k \vn k}$ and $\Omega^{\hat{\vn m}\vn k}$. Taking into account the above parametrization of the magnetization direction by $\theta$, we were thus able to access efficiently the anomalous Hall conductivity $\sigma_{ij}$, the torkance $\tau_{yj}$, and the spiralization $D_{yj}$:
\begin{align}
\sigma_{ij} &= \frac{e^2}{h}\frac{1}{2\pi} \int 2\text{Im} \sum\limits_{n}^{\text{occ}} \bigg\langle \frac{\partial u_{\vn k n}}{\partial k_i} \bigg | \frac{\partial u_{\vn k n}}{\partial k_j}\bigg\rangle dk_x dk_y \, , \label{eq:ahc} \\
 \tau_{yj} &= e \int 2\text{Im} \sum\limits_{n}^{\text{occ}} \bigg\langle \frac{\partial u_{\vn k n}}{\partial \theta} \bigg | \frac{\partial u_{\vn k n}}{\partial k_j}\bigg\rangle dk_x dk_y \, , \label{eq:sot} \\
 D_{yj} &= \frac{1}{V} \int \text{Im} \sum\limits_{n}^{\text{occ}} \bigg\langle \frac{\partial u_{\vn k n}}{\partial \theta} \bigg | h_{\vn k n} \bigg | \frac{\partial u_{\vn k n}}{\partial k_j}\bigg\rangle dk_x dk_y \, , \label{eq:dmi}
\end{align}
with the same definitions as in the main text.
Convergence of these quantities was achieved using $1024\times 1024$ $\vn k$-points in the Brillouin zone. We obtained the mixed Chern number $\mathcal Z(k_y)=1/(2\pi)\int 2\text{Im}\sum_{n}^{\text{occ}} \langle \partial_{\theta} u_{\vn k n} | \partial_{k_x} u_{\vn k n}\rangle d\theta dk_x$ by integrating the mixed Berry curvature on a uniform mesh of 1024 $k_x$-values and $512$ angles $\theta$ in $[0,2\pi)$.

\vspace{0.7cm}
\noindent{\bf Supplementary Note 3 | Anisotropy with magnetization direction in semi-hydrogenated Bi bilayer}
\\In Supplementary Fig.~1, we show the dependence of anomalous Hall conductivity, torkance, and spiralization on the magnetization direction $\hat{\vn m}=(\sin\theta,0,\cos\theta)$ in the semi-hydrogenated bismuth film. For general magnetization directions, both torkance and spiralization display also small non-zero components $\tau_{yy}$ and $D_{yy}$, respectively, since the shape of these response tensors is dictated by the crystal symmetries and not due to Onsager's reciprocity relations. When the Weyl point emerges in the electronic structure at $\theta=43^\circ$, the system undergoes a topological phase transition from a Chern insulator to a trivial magnetic insulator, which is accompanied by a jump in $\sigma_{xy}$ and a similar drop of $\tau_{ij}$. As apparent from Supplementary Figs.~1 and~2, the torkance $\tau_{yx}$ is still remarkably prominent in the regime of the trivial insulator, i.e., for $\theta>43^\circ$. 

\vspace{0.7cm}
\noindent{\bf Supplementary Note 4 | GaBi film with exchange field}
\\Using the \fleur{} code~\cite{fleur}, we performed self-consistent electronic structure calculations of an intrinsically non-magnetic GaBi film, employing the generalized gradient approximation and a plane-wave cut-off of $4.0\,a_0^{-1}$, where $a_0$ is Bohr's radius. The in-plane lattice constant was $8.5\,a_0$ and we chose a muffin tin radius of $2.45\,a_0$ for both atom species. Subsequently, we constructed $16$ maximally-localized Wannier functions out of $32$ energy bands with the frozen window extending up to $2\,$eV above the Fermi level. Finally, in order to substantiate the predicted effect of monopole-driven spin-orbit torques, the exchange term $\vn B\cdot \vn \sigma$ was added to the corresponding tight-binding Hamiltonian, where $\vn \sigma$ is the vector of Pauli matrices and $\vn B=B_0(\sin\theta,0,\cos\theta)$ denotes the imposed exchange field.

\vspace{0.7cm}
%\newpage
\noindent{\bf Supplementary Figures}
\begin{figure}[h] %[b]
\centering
\includegraphics{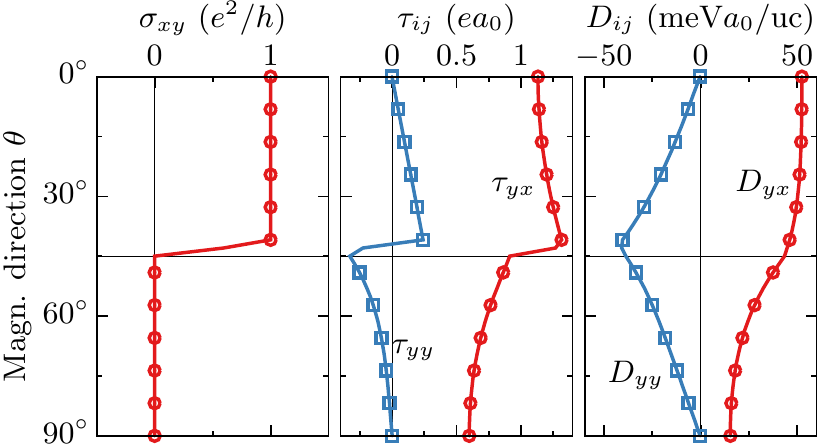}
\caption{{\bf Dependence on magnetization direction in semi-hydrogenated Bi bilayer.} The anomalous Hall conductivity $\sigma_{xy}$, the torkance $\tau_{ij}$, and the spiralization $D_{ij}$ at the actual Fermi level as a function of the magnetization direction $\hat{\vn m}=(\sin\theta,0,\cos\theta)$. A mixed Weyl point emerges in the electronic structure at $\theta=43^\circ$.}
\label{fig:anisotropy}
\end{figure}

\begin{figure}[h] %[b]
\centering
\includegraphics{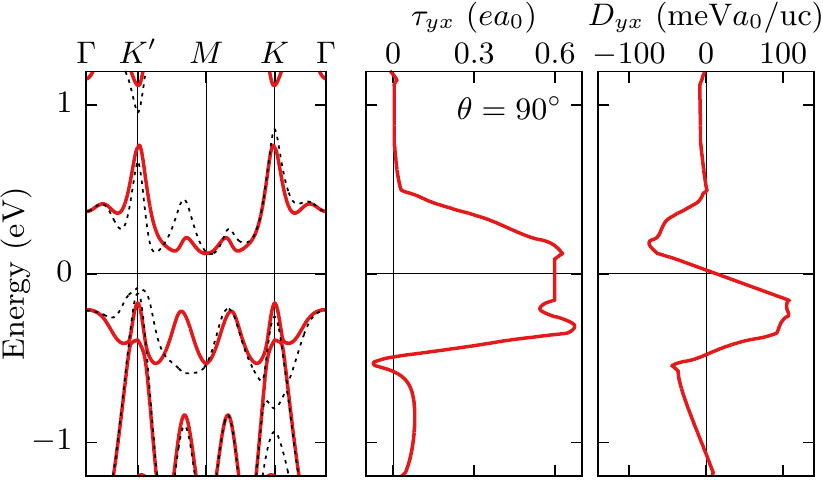}
\caption{{\bf Semi-hydrogenated Bi bilayer with in-plane magnetization.} Band structure, torkance $\tau_{yx}$, and spiralization $D_{yx}$ as function of the position of the Fermi level if the magnetization is parallel to the film, i.e., $\theta=90^\circ$. For comparison, dashed lines in the left panel denote the spectrum for an out-of-plane magnetization ($\theta=0^\circ$).}
\label{fig:anisotropy2}
\end{figure}

\end{document}